\documentclass[11pt,twocolumn,twoside]{osajnl}
\usepackage{mathtools}
\usepackage{amsmath}
\usepackage{units}
\usepackage{upgreek}
\graphicspath{ {./image/} }

\journal{ao} 

\setboolean{shortarticle}{true}

\title{Deep learning-driven adaptive optics for laser wavefront correction}
\author[1,2,*]{Jikai Wang}
\author[1]{Sven Burckhard}
\author[1]{Sonam Smitha Ravi}
\author[3]{Dominik Bauer}
\author[1]{Volker Rominger}
\author[2,4]{Stefan Nolte}
\author[1]{Daniel Flamm}
\affil[1]{TRUMPF Laser-und Systemtechnik SE, Johann-Maus-Strasse 2, 71254 Ditzingen, Germany}
\affil[2]{Friedrich Schiller University Jena, Institute of Applied Physics, Abbe Center of Photonics, Albert-Einstrein-Str.~15, 07745 Jena, Germany }
\affil[3]{TRUMPF Laser SE, Aichhalder Str.~39, 78713 Schramberg, Germany}
\affil[4]{Fraunhofer Institute for Applied Optics and Precision Engineering, Albert-Einstein-Str.~7, 07745 Jena, Germany}
\affil[*]{jikai.wang@uni-jena.de}

\ociscodes{(140.3295) Laser beam characterization, (140.3300) Laser beam shaping, (140.3390) Laser materials processing.}
\doi{\url{https://doi.org/xxxxx/AO.xxxxx}}

\begin{abstract}
\textcolor{black}{We report on an intensity-only and deep-learning based method for laser beam characterization that allows to predict the underlying optical field within milliseconds. A simple near-field / far-field camera setup enables online control of an adaptive optics to optimize beam quality. The robustness and precision of the method is enhanced by applying the concept of phase diversity based on spiral phase plates.}
\end{abstract}

\setboolean{displaycopyright}{false}

\begin{document}

\maketitle
J.~Wang \textit{et al.}, Appl.~Opt. \textbf{XX}, XXXX (2025).\vspace{0.1cm }\\ 
\textbf{© 2025 Optica Publishing Group. One print or electronic copy may be made for personal use only. Systematic reproduction and distribution, duplication of any material in this paper for a fee or for commercial purposes, or modifications of the content of this paper are prohibited.}
\section{Introduction}\label{section:1}

The generation of the highest laser intensities is known to require pulses with highest energies and shortest durations as well as the ability to focus them as close to the diffraction limit as possible \cite{stark2023pulses, pfaff2023nonlinear, dominik2022thin}. The latter specification depends on the laser beam quality originally provided by the source and the performance of the optical systems required for beam guidance, shaping and focusing \cite{wang2024phase}. Almost without exception, the beam propagation ratio, or $M^2$-parameter \cite{siegman1998maybe}, is used to quantify achieved focus qualities, see, e.g.~Refs.~\cite{stark2023pulses, pfaff2023nonlinear, dominik2022thin}. The use of a standardized measurement procedure, see ISO 11146-1/2/3 \cite{ISO11146} ensures comparability of different laser architectures and focusing situations. Comparability, however, comes at the expense of losing
information as complex optical fields are described with a single parameter only \cite{siegman1998maybe}. However, recently, we have demonstrated that the standardized measurement procedure \cite{ISO11146} combined with a phase retrieval approach \cite{allen2001phase} allows access to complex phase distributions with an accuracy down to $\lambda/25$ (RMSE) \cite{wang2024phase}. A camera-based, intensity-only metrology is, thus, able to reconstruct complex optical fields with highest fidelity. There is one disadvantage remaining: the required caustic measurement is in the order of one minute---far too long to investigate and compensate for dynamic processes.

In this work, a machine learning algorithm based concept is introduced that allows for optical field prediction from intensity-only signals in real-time. A convolutional U-Net \cite{ronneberger2015u} is trained from image mapping between phase profiles generated by a liquid-crystal-based spatial light modulator (SLM) and corresponding measured intensity profiles in the near- and far-field. We make use of spiral phase diversity \cite{echeverri2016Vortex} to enhance the quality of the phase prediction where a vortex phase distortion supports the detection of smallest phase aberrations in intensity features. In contrast to existing work, see, e.g. Long \textit{et al.~}\cite{long2024situ}, our approach relies on (at least) two camera measurements in a far-field and near-field-like arrangement. This enhances robustness of the metrology and enables an unambiguous phase prediction. By intention, the strategy followed in this work is to predict optical fields not modal based but zonal \cite{ISO15367}. The predicted output for our training and compensation is, thus, not available in terms of mode coefficients, but directly as a two-dimensional phase distribution $\theta\left(x,y\right)$. This approach increases the numerical effort, but enables the prediction of phase profiles with jumps or singularities that can only be poorly described with continuous basis functions such as Zernike modes \cite{noll1976zernike}---which we use nonetheless in our fundamental study, cf.~Sec.~\ref{sec:OF}. In addition, the present work provides a thorough tolerance analysis where the physical limitations of machine learning-based optical field analysis are presented, cf.~Sec.~\ref{simu}. Finally, we apply our approach to experimental data representing single-mode-like cases with beam qualities ``close'' to the diffraction limit $M^2 \in \left[1 \dots 2\right]$. By means of this data the benefit of our approach for beam quality analysis of high power or high energy lasers \cite{sutter2019high} is discussed. Here, the associated phase profiles are often influenced by thermal lenses within optical components caused by high average powers \cite{mansell2001evaluating} or by their $B$-integral resulting from the highest intensities \cite{perevezentsev2007comparison}, cf.~Sec.~\ref{AOsetup}.

\section{Fundamentals}
\label{sec:OF}
The following sections provide the optical fundamentals for field manipulation with phase-only transmission functions (Sec.~\ref{sec:OF}\,\ref{sec:AOV}) as well as the introduction of the machine learning algorithm (Sec.~\ref{sec:OF}\,\ref{sec:U_net}).
\subsection{Phase modulation and phase diversity}
\label{sec:AOV}
In our concept, the input radiation changes in two ways. On the one hand, 
we apply a variety of phase modulations based on Zernike modes ($T_{\text{mod}}$) to train our network. On the other hand, we use the concept of phase diversity to facilitate the detection of induced phase perturbations ($T_{\text{vort}}$). The linearly polarized (scalar) optical field $E_{\text{in}}\left(x,y\right)$ is modulated by two phase-only transmission functions $E_{\text{out}}\left(x,y\right) = E_{\text{in}}\left(x,y\right)T_{\text{mod}}\left(x,y\right)T_{\text{vort}}\left(x,y\right)$ with
\begin{equation}
\begin{split}
\label{eq:Zern}
    T_{\text{mod}}\left(x,y\right) &= \exp{\left[\imath \phi_{\text{mod}}\left(x,y\right)\right]} \\ &=  \exp{\left[\imath \uppi \sum_{mn}c_{mn} Z_{mn}\left(x,y\right) \right]}
\end{split}
\end{equation}
and
\begin{equation}
\begin{split}
\label{eq:Vort}
    T_{\text{vort}}\left(\theta\right) = \exp{\left(\imath \ell \theta \right)}.
\end{split}
\end{equation}

In Eq.~(\ref{eq:Zern}) a set of Zernike modes $\left\{Z_{mn}\left(x, y\right)\right\}$ with $n$-th radial and $m$-th azimuthal order \cite{noll1976zernike} determine the phase distortions $T_{\text{mod}}$ with corresponding real-valued mode coefficients $c_{mn}$ which are applied for training the network. For our experiments, this set is composed of the first $15$ Zernike modes excluding the piston and both tilts $c_{\left(0,0\right)} = c_{\left(1,1\right)} = c_{\left(-1,1\right)} = 0$, as they do not impact the beam quality. We would like to emphasize that although a set of Zernike modes is used in this fundamental study, we are not limited to this---especially not to a modal basis. Any other method providing a zonal set of phase modulations can be applied.

The stationary transmission function $T_{\text{vort}}$ defined by Eq.~(\ref{eq:Vort}) with the topological charge $\ell = \text{const.}$ is used to enhance the diversity in analyzing phase distortion with intensity-only signals \cite{pang2020focal}. Here, $E_{\text{in}}$ is transformed into an orbital angular-momentum-carrying beam \cite{padgett2015divergence}. Under ideal conditions, in the absence of phase disturbances, a ring profile is formed in the far field whose optical field is very similar to that of a Laguerre-Gaussian mode \cite{saghafi1998beam}. Deviations from this ideal case result in easily recognizable intensity features, such as asymmetries or inhomogeneities \cite{ohland2019study}.

Phase disturbances naturally appear in the intensity profile during the propagation of the radiation in free space or in the transition to the far field\cite{wang2024phase}. However, changes in intensity can be difficult to detect if they are lost in the noise of the detector or if the associated dynamic range is insufficient. Therefore, we ``force'' the radiation to show an additional property---the diffraction at the spiral phase transmission $T_{\text{vort}}$. We will demonstrate the benefits of applying $T_{\text{vort}}$ to the test radiation in Secs.~\ref{simu} and \ref{AOsetup}. However, we do not claim this to be the optimized approach for phase diversity. Alternatives may be found by using correlation filters \cite{flamm2012mode}, axicon holograms \cite{leach2006generation}, triangular apertures \cite{hickmann2010unveiling} or combinations thereof.

\begin{figure*}
\centering
\includegraphics[width=0.7\textwidth]{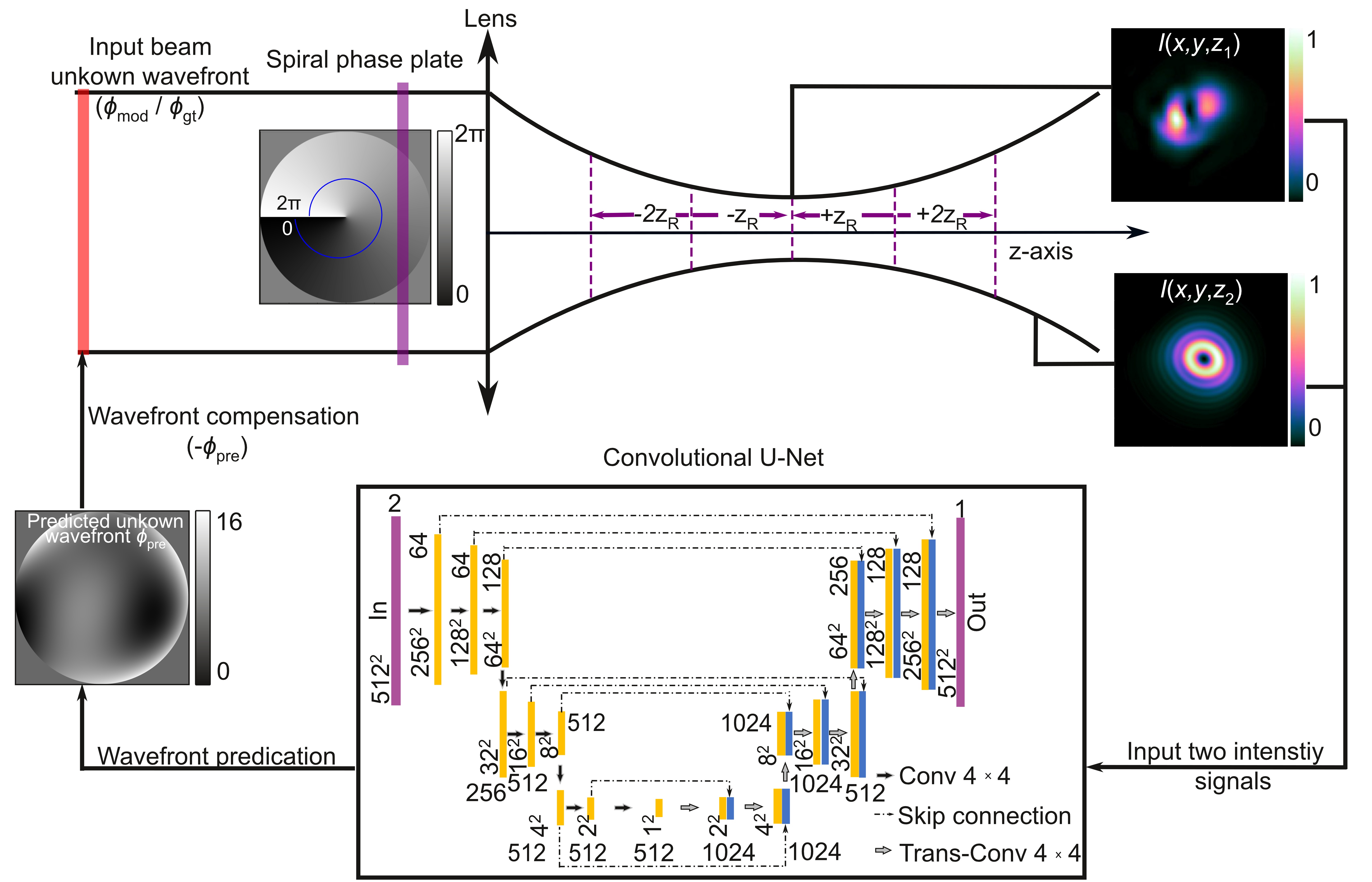}
\caption{Adaptive optics with phase diversity based on Convolutional U-Net: the input beam $E_{\text{in}}$ with unknown phase $\phi_{\text{mod}}$ is propagating through a spiral phase plate and focused with a lens (top). The far field $I_1$ with ($ -z_{R}< z < +z_{R}$) and the near field  $I_2$ with  ($ z<-2z_{R}$, $z>+2z_{R}$) are recorded to predict the input unknown wavefront $\phi_{\text{mod}}$ using U-Net algorithm, followed by compensation with $-\phi_{\text{pre}}$. Convolutional U-Net architecture consists of $9$ convolutional layers and $9$ transposed convolutional layers (bottom). The input is a grayscale near and far field intensity image pair of size ($512 \times 512 \times 2$).The $\phi_{\text{pre}}$ is the result predicted from the convolutional U-Net.}
\label{fig:main_principle}
\end{figure*}

The adaptive optical setup for training and wavefront compensation is shown schematically in Fig.~\ref{fig:main_principle} (top). Here, the input beam $E_{\text{in}}$ is phase manipulated using $T_{\text{zern}}$ and $T_{\text{vort}}$, respectively, and is transformed to the far field using a lens with the corresponding intensity $I\left(x,y,z=z_1\right) \eqqcolon I_1$. The $z$-axis equals the propagation direction and position $z_1$ is located within one Rayleigh length $z_\text{R}$ around the waist $\left( -z_\text{R} < z_1 < +z_\text{R} \right)$. 
As mentioned in Sec.~\ref{section:1} and in contrast to existing approaches \cite{long2024situ}, a second near-field-like intensity distribution $I\left(x,y,z=z_2\right) \eqqcolon I_2$ is required with $\left( z_2 < -2z_{\text{R}} \right)$ or $\left( z_2 > +2z_{\text{R}}\right)$. This near-field-like signal located beyond two Rayleigh lengths from the waist ensures the sign-correct determination of the phase distortion at hand. With our phase diversity approach, where the sign of the spiral phase transmission $T_{\text{vort}}$, cf.~Eq.~(\ref{eq:Vort}), is well defined, even the sign of phase singularities can be predicted unambiguously.

The need for a second intensity signal increases the optical and numerical effort. However, in order to measure several intensity signals in different propagation steps simultaneously, clever optical solutions have been proposed in the literature \cite{scaggs2012real}. For the sake of simplicity, we will work with two cameras in the following.

\subsection{Machine learning algorithm: Convolutional U-Net}
\label{sec:U_net}
The U-Net is an architecture originally introduced by Ronneberger \textit{et al.}~\cite{ronneberger2015u}. Our adaptations are shown in Fig.~\ref{fig:main_principle} (bottom), which is inspired by image-to-image translation with conditional adversarial networks proposed by Isola \textit{et al.}~\cite{isola2017image} \cite{wang2021deep}. The input stack that consists of the pair of near- and far-field intensity distributions $I_1$, $I_2$, is of size $\left(512 \times 512 \times 2\right)$ after preprocessing, including normalization, background subtraction, and cropping \cite{sonka1993image}. As output a phase profile of size $\left(512 \times 512 \times 1\right)$ is provided; see Fig.~\ref{fig:main_principle} (bottom). The network consists of a convolutional encoder \cite{sonka1993image} with $9$ convolutional layers and a transposed convolutional decoder \cite{isola2017image}  with $9$ convolutional layers. In our case, a $\left(4 \times 4\right)$ kernel \cite{ding2022scaling} is used to train U-Net. As proposed by Isola \textit{ et al.}, skip connections \cite{isola2017image} are also applied within the U-Net, represented by the dashed lines in Fig.~\ref{fig:main_principle}. This aims to keep low-level features, such as the structure of phase profiles, at a deep layer of the encoder \cite{ding2022scaling}. As a loss function used for training and evaluation, we introduce the mean square error (MSE) that quantifies the differences in the predicted phase distribution $\phi_{\text{pre}}\left(x,y\right)$ with respect to the ground truth $\phi_{\text{gt}}\left(x,y\right)$. 

\begin{equation}
\label{eq:MSE}
 \text{MSE}=\frac{\int_{x_1}^{x_2}\int_{y_1}^{y_2} \left[\phi_{\text{pre}}(x,y) - \phi_{\text{gt}}(x,y) \right]^2 \,\text{d}x \text{d}y}{\left(x_2-x_1\right)\left(y_2-y_1\right)}.
\end{equation}
Additionally, in order to express mentioned differences in terms of the wavelength $\lambda$ we use the root-mean-square error (RMSE) with wavelength normalization
\begin{equation}
\label{eq:rmse}
    \delta = \sqrt{\text{MSE}} \times \lambda/2\uppi.
\end{equation}
\textcolor{black}{Alternative evaluation indicators for phase errors could be derived indirectly from corresponding (focus) intensities. The Strehl ratio \cite{mahajan1982strehl}, which is often used to evaluate the performance of adaptive optics, could be applied here, too. However, since we have direct access to the reference phase $\phi_{\text{gt}}\left(x,y\right)$, realized with the liquid crystal display, cf.~Sec.~\ref{AOsetup}~\ref{setup}, the phase error $\delta$ is used.}

Equations (\ref{eq:MSE}) and (\ref{eq:rmse}) require $\phi_{\text{pre}}$ which is the result of our U-Net operation denoted by function $F$

\begin{equation}
\label{eq:network}
 \phi_{\text{pre}}\left(x,y\right)=F\left[I_{1}\left(x,y\right), I_{2}\left(x,y\right), w \right]. 
\end{equation}
The parameter $w$ refers to the optimized weights connecting the network layers, see Fig.~\ref{fig:main_principle} (bottom).
The network implemented on TensorFlow 2.11.0 and Python Version 3.10.11 on a GeForce RTX 4090 GPU requires a training time of $\sim \unit[300]{mins}$. The prediction effort is $\sim \unit[17]{ms}$ on a standard laptop. The training data is composed of a set of 20,000 pairs of near- and far-field intensity profiles, with training data (70\%), test data (15\%) and validation data (15\%).

\section{Synthetic Data}
\label{simu}
Before testing our approach in experiments, we apply it to synthetically generated data sets to optimize image processing steps, test neural network robustness (Sec.~\ref{simu}\,\ref{sec:imageproc}) and determine the limits of the method (Sec.~\ref{simu}\,\ref{sec:training}).

\subsection{Data preparation and robustness analysis}\label{sec:imageproc}
U-Net input image signals are typically scaled to smaller sizes to reduce computational efforts \cite{beheshti2020squeeze}. However, various interpolation steps have to be carried out in order to determine the resolution of the spatial light modulator---finally used for phase compensation. We found Inter area \cite{parihar2022comprehensive} and Lanczos \cite{turkowski1990filters} to be effective interpolation methods. The original phase mask of $\left(512 \times 512 \right)$ size is downsampled to $\left(64 \times 64\right)$ and upsampled again back to $\left(512 \times 512\right)$. Resulting overlap integrals \cite{albert2014laser} between the original and interpolated optical fields exceed $99\%$, demonstrating the effectiveness of these interpolation techniques in preserving phase profiles during scaling. 

\begin{figure*}
\centering
\includegraphics[width=0.6\textwidth]{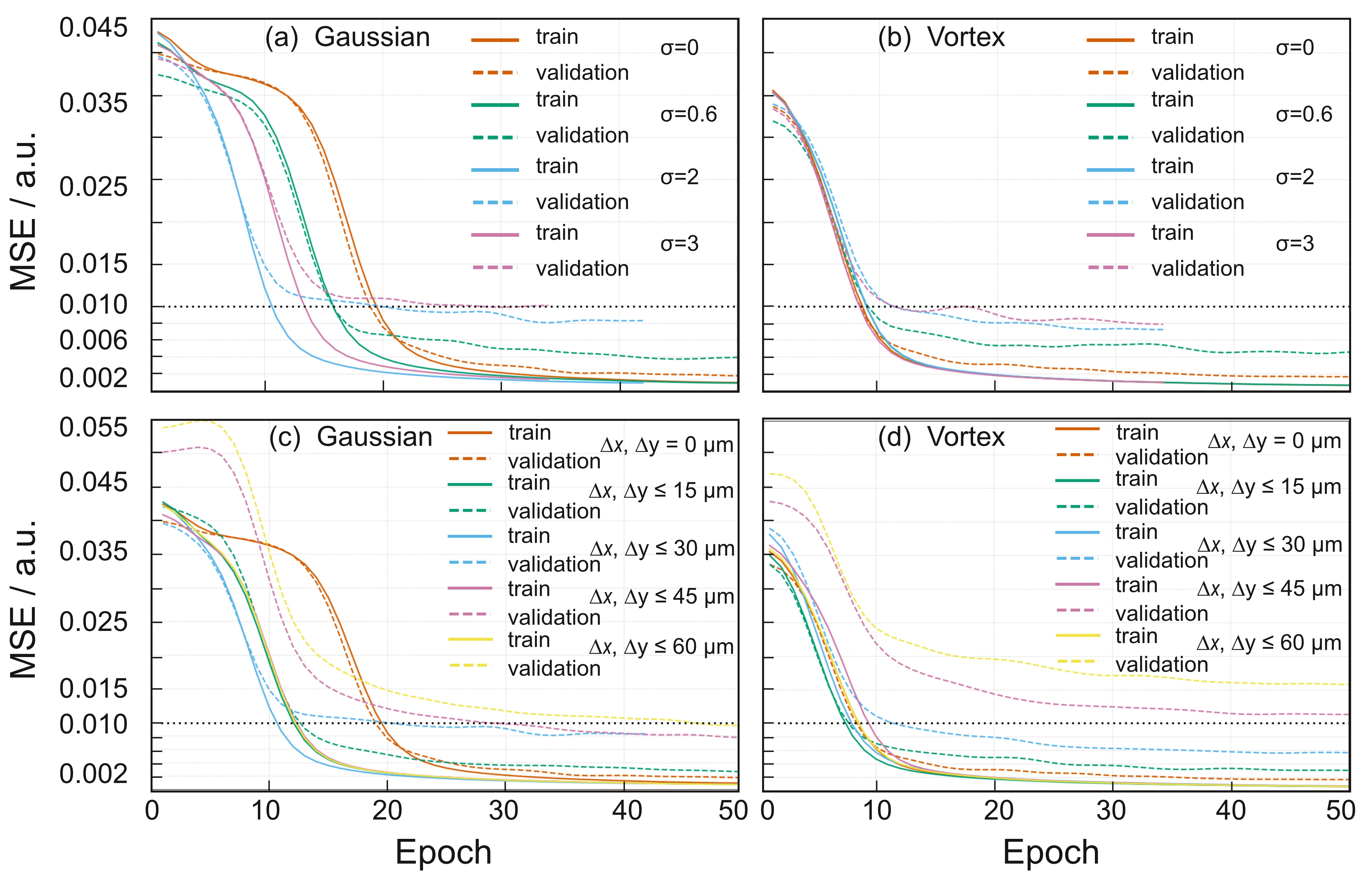}
\caption{Neural network's robustness analysis from synthetic data by adding noise (a), (b) and transverse (camera) shifts (c),  (d) without (``Gaussian'') and with spiral phase diversity (''Vortex''). In all cases the MSE parameter, cf.~Eq.~(\ref{eq:MSE}), is used to evaluate the U-Net performance as a function of number of epochs.}
\label{fig:sensitivity_analysis}
\end{figure*}

Originally, we expected a low network robustness under noisy input and unstable optical conditions. However, it was found that high-quality operation is achieved even in challenging noise scenarios. To test this, we add three Gaussian noise levels \cite{gonzalez2009digital} ($\sigma = 0.6$, $\sigma = 2$ and $\sigma = 3$) to the input data, as shown in Fig.~\ref{fig:sensitivity_analysis} where the U-Net performance can be seen without ``Gaussian'' (a) and with spiral phase diversity ``Vortex'' (b). The standard deviation $\sigma=0.6$ matches the real-world noise level with the applied camera. Higher noise levels represent more extreme scenarios to evaluate our method's limits. Training results converge to $\text{MSE}<0.01$ ($\sim \lambda/63$, ``Gaussian'') and $\text{MSE}<0.008$ ($\sim \lambda/71$, ``Vortex''), and, thus, behave similarly, but converge slightly faster in the Vortex case ($<10$ epochs). In any case, we have observed a high degree of robustness even at high noise levels.

A further experimental influence with an impact on the training performance is beam position stability modeled as transverse shifts $\Delta x$, $\Delta y$ of the near- and far-field intensity signals $I_{1, 2}\left(x+\Delta x, y+\Delta y\right)$. Here, we assume
$\Delta x, \Delta y \in \left[ \unit[-60]{\upmu m} \dots \unit[60]{\upmu m} \right]$ which corresponds to relative shift of $\approx 1\%$ of the raw beam diameter. The training prediction performance shown in Fig.~\ref{fig:sensitivity_analysis}, again, demonstrates a similar behavior for both cases without (c) and with spiral phase diversity (d), reaching MSE values around 0.01 ($\sim \lambda/63$) with stronger fluctuations in the Gaussian case especially for small number of epochs ($< 10$). 
This section's robustness analysis has shown that camera noise and the spatial stability of our metrology have a small impact to the network's prediction performance. This could be due to the applied interpolation algorithms required for down- and upsampling, respectively, acting as low-pass filter.

\subsection{Training results with synthetic datasets}\label{sec:training}
The network's training performance without (``Gaussian'') and with spiral phase diversity (``Vortex'') shown in Fig.~\ref{fig:sensitivity_analysis}, demonstrates the ability of the U-net model to converge. The noise-free validation curve reaches an MSE of $\sim 2\text{E-}3$ after 50 epochs, which corresponds to a phase prediction accuracy of $\delta \approx \lambda/140$. This performance results from using a combination of near- and far-field inputs. The prediction accuracy was also tested with near-field-like and far-field data separately. Our previous work \cite{wang2024laser} showed that combining both datasets reduces MSE compared to using either alone. Additionally, the approach enables a sign-correct phase prediction as the focused $I_1$ and defocused $I_2$ measured intensities also cover the propagation behavior, cf.~Sec.\,\ref{sec:OF} \ref{sec:AOV}. 

\begin{figure*}[h]
 \centering
\includegraphics[width=0.6\textwidth]{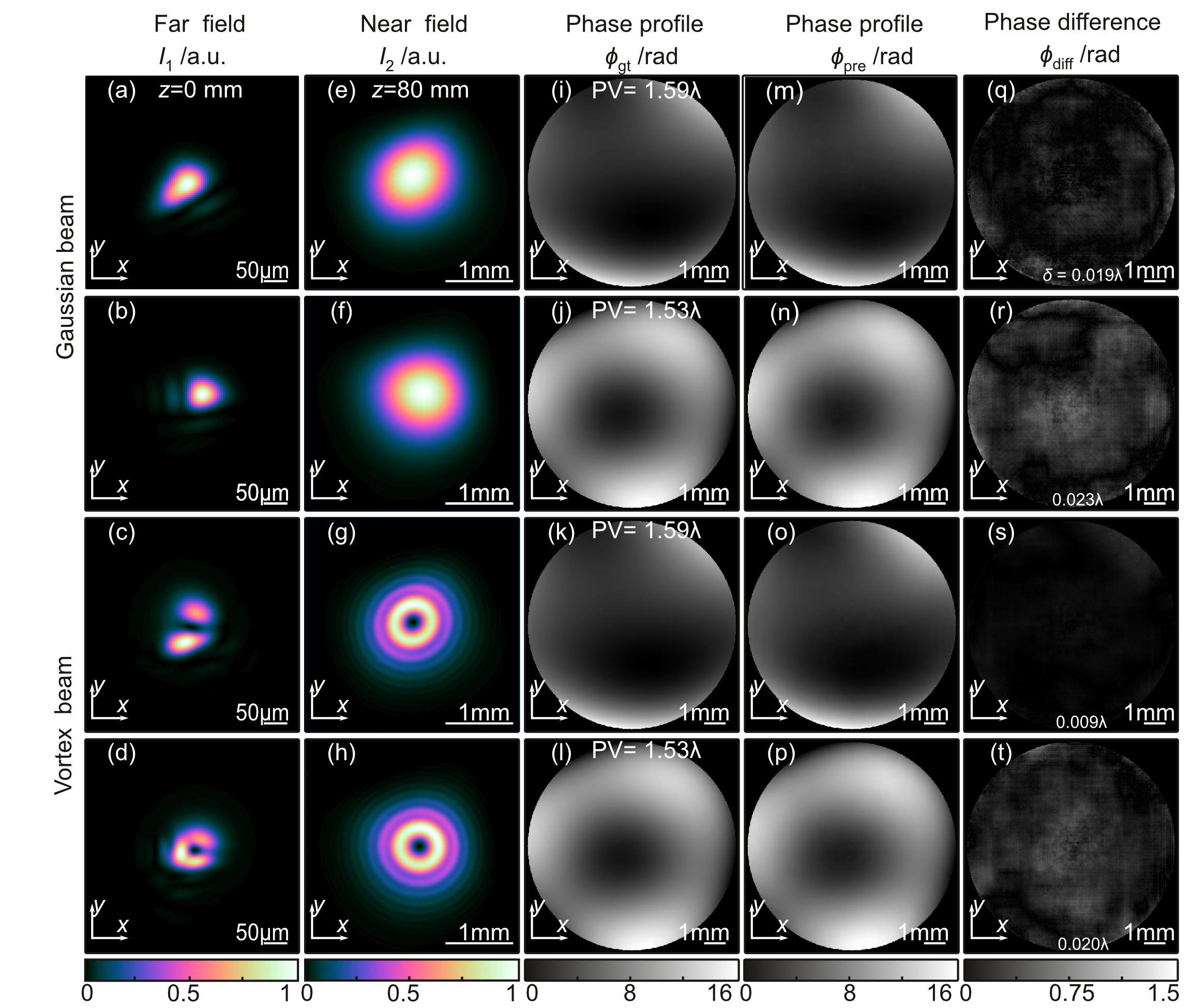}
\caption{Phase profile prediction result: far field $I_{1}$, (a) -- (d) and near field $I_{2}$, (e) -- (h), located at the focus $z = \unit[0]{mm}$ (far-field) and $z = \unit[80]{mm}$ (near-field-like), respectively. The ground truth phase profiles $\phi_{\text{gr}}$ (i) -- (l) are compared with the predicted phase profiles $\phi_{\text{pre}}$ (m) -- (p). Their absolute phase difference $\phi_{\text{diff}}$ (q) -- (r) shows that the average $\delta$ using spiral phase diversity (``Vortex'') is decreased by $\sim 32\%$ in comparison to the standard case (``Gaussian'').}
\label{fig:Gaussian_sim}
\end{figure*}

We investigate two examples of phase profiles predicted from the convolutional U-net using the two intensity signals. For both cases we compare the U-net performance without (``Gaussian'') and with spiral phase diversity (``Vortex''), see Fig.~\ref{fig:Gaussian_sim}. Here, we use the same ground truth phase $\phi_{\text{gt}}$ as shown in Fig.~\ref{fig:Gaussian_sim}(i)–(l). The peak-to-valley (PV) values are $1.59\lambda$ and $1.53\lambda$, respectively. The far-field and near-field-like signals ($I_{1}$, $I_{2}$) are displayed in Fig.~\ref{fig:Gaussian_sim}\,(a)\,--\,(d) and (e)\,--\,(h), respectively. They are generated from an ideal Gaussian beam of diameter $d=\unit[6]{mm}$, wavelength ${\lambda}=\unit[1030]{nm}$, and a lens with focal length $f=\unit[300]{nm}$.  The predicted phase profiles $\phi_{\text{pre}}$ are shown in Fig.~\ref{fig:Gaussian_sim}\,(m)\,--\,(p) and the prediction accuracy in (q)\,--\,(t) is reflected with the absolute phase difference $\left(\phi_{\text{diff}}=|\phi_{\text{gt}} - \phi_{\text{pre}}|\right)$ between the ground truth and the predicted phase profiles. To quantify the phase difference, the root-mean-square error $\delta$ is used as defined in Eq.~(\ref{eq:rmse}) revealing a higher accuracy in phase prediction when spiral phase diversity is applied [$\delta$-reduction of $53\,\%$ comparing (q) and (s) and $13\,\%$ comparing (r) and (t)].

These synthetic training results demonstrate the network's ability to accurately predict phase profiles from two intensity signals $I_{1}$ and $I_{2}$. Furthermore, the predicted phase profiles exhibit strong statistical agreement with the ground truth, evidenced by RMSE differences down to $\delta \approx \lambda/140$. Thus, the theoretically achievable discrepancies are so small that they have no technical-optical relevance.

\section{Experimental validation}
\label{AOsetup}
\textcolor{black}{This section presents the adaptive optical experimental setup (Sec.~\ref{AOsetup}\,\ref{setup}). The training results using experimental data sets are discussed in Sec.~\ref{AOsetup}\,\ref{training:exp}. In Sec.~\ref{AOsetup}\,\ref{verification}, random aberrated phase masks are tested for wavefront compensation, demonstrating beam quality enhancement.} 

\subsection{Experimental setup}
\label{setup}
\begin{figure}[t]
    \centering
    \includegraphics[width=0.4\textwidth]{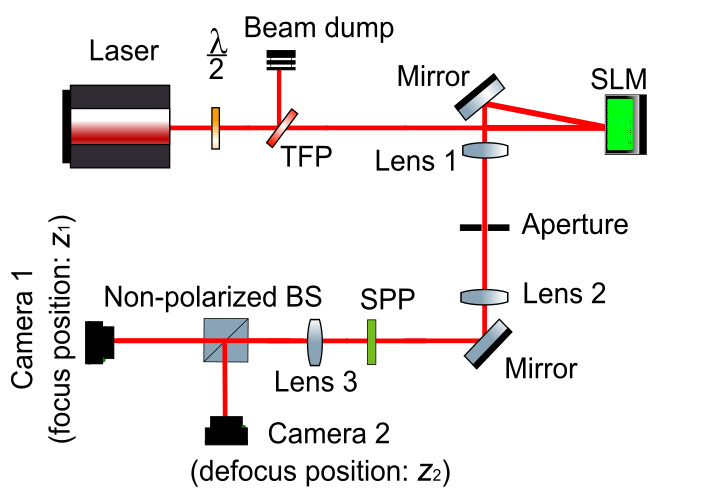}
    \caption {Experimental setup: Thin film polarizer (TFP), with half wave-plate (HWP) to attenuate the beam power. Liquid-crystal-on-silicon-based spatial light modulator (SLM). Imaging setup ($4f$-like) consisting of Lens 1 and 2. Aperture for blocking 0 order beam. SPP for vortex beam generation. Lens 3 for generating the far and near fields recorded by camera 1 and camera 2 \textcolor{black}{ at positions $z_1$ and $z_2$, respectively, see Sec.~\ref{sec:OF}~\ref{sec:AOV}.}}
    \label{fig:experiment Setup_1}
\end{figure}

For experiments conducted in this section, the optical setup shown in Fig.~\ref{fig:experiment Setup_1} is used. A low-power fiber laser average power $P=\unit[24]{mW}$, a beam diameter of $d_0=\unit[6]{mm}$, a pulse duration of ${\tau_{p}}=\unit[120]{ps}$, operating at ${\lambda}=\unit[1030]{nm}$ serves as a light source. A half-wave plate in combination with a thin film polarizer (TFP) allows us to select suitable intensities for our near- and far-field measurements, being collected with camera 1 and camera 2 (IDS UI-3370CP-NIR-GL) and being located at $z_1$, $z_2$, \textcolor{black}{see Sec.~\ref{sec:OF}~\ref{sec:AOV}}. The liquid crystal-based spatial light modulator (SLM, Hamamatsu X15223 series) operates in phase-only mode, displaying the phase profiles (aberrations), resulting in altered intensity profiles at $z_1 = \unit[0]{mm}$ and $z_2 = \unit[80]{mm}$ after lens 3 ($f=\unit[300]{mm}$). The spiral phase plate SPP (fabricated by Vortex Photonics ) placed before lens 3 applies a topological charge $\ell = 1$ to the input optical field.

\subsection{Training results with experimental datasets}
\label{training:exp}
\begin{figure}
\centering\includegraphics[width=\columnwidth]{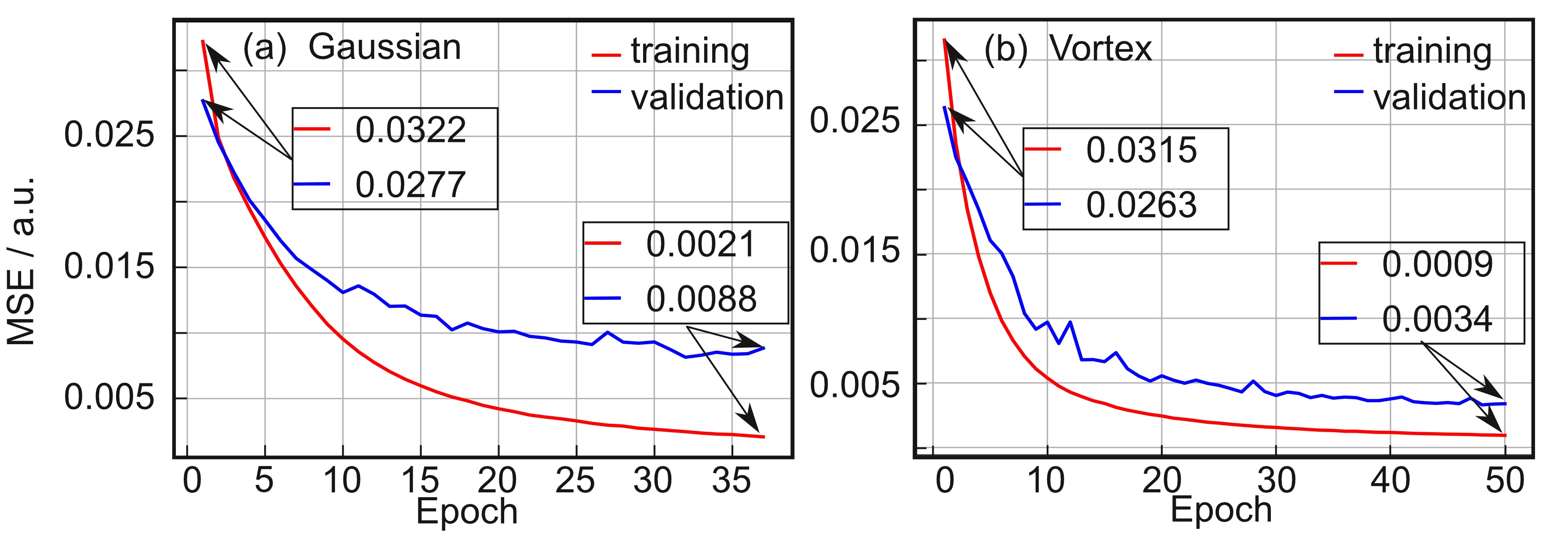}
\caption{Training (blue) and validation (red) loss curves in terms of MSE parameter without ``Gaussian'' (a) and with spiral phase diversity ``Vortex'' (b) as a function of epochs. The validation MSE reaches $\sim$ 5E$-$3 with spiral phase diversity, compared to $\sim$ 8E$-$3 for the Gaussian case.}
\label{fig:loss function}
\end{figure}

Training and validation results based on experimental data sets (20,000 pairs of $I_1$ and $I_2$) are depicted in Fig.~\ref{fig:loss function}. Again, we compare the performance without (a) (``Gaussian'') and with spiral phase diversity (b) (``Vortex''). Both training curves (red) converge to the experimentally achievable limit which amount to $\text{MSE} \approx 2\text{E-}3$  ($\delta \approx \lambda/140$) and $\text{MSE} \approx 1\text{E-}3$  ($\delta \approx \lambda/200$), respectively. Thus, the benefit of spiral phase diversity can already seen from the training data with an improvement of $\sim 43\,\%$.

The validation curves (blue) follow this convergence behavior. Statistically, we achieve phase differences to the ground truth of $\text{MSE}\approx 9\text{E-}3$ ($\delta \approx \lambda/33$) (``Gaussian'') and $\text{MSE} \approx 4\text{E-}3$ ($\delta \approx \lambda/99$) (``Vortex''). Applying spiral phase diversity enhances the prediction accuracy by $\sim 55\,\%$. Additionally, we observe a faster convergence speed for the ``Vortex'' case. Already after 20 epochs the Gaussian-MSE-limit is achieved where 35 epochs are required. This improvement is primarily attributed to the network's ability to extract more features from vortex beams, facilitated by their inherent phase diversity, compared to Gaussian beams \cite{huang2021all}.

\begin{figure*}[h]
\centering\includegraphics[width=0.6\textwidth]{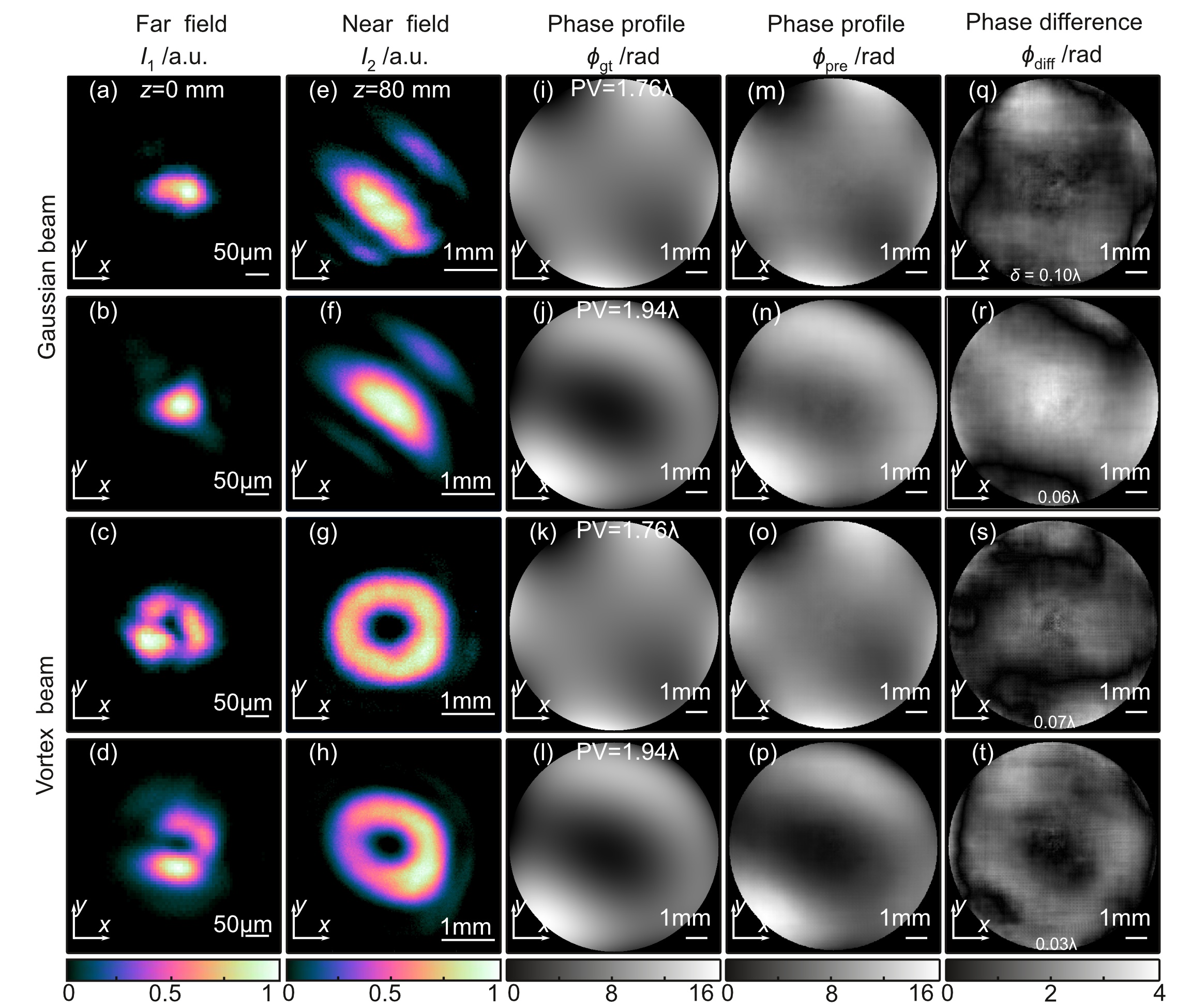}
\caption{Phase profile prediction from experimental data. Measured far-field (a) -- (d) and near-field-like intensity signal (e) -- (h) without (``Gaussian'') and without (``Vortex'') spiral phase diversity. The ground truth phase profiles $\phi_{\text{gr}}$ (i) -- (l), are compared with the predicted phase profiles $\phi_{\text{pre}}$ (m) -- (p). The phase differences (q) -- (t) show a more accurate phase prediction when spiral phase diversity is applied, as evidenced by the RMSE-parameter, e.g. a reduction from $\delta = 0.06\lambda$ ((r), ``Gaussian'') to $\delta = 0.03\lambda$ ((s), ``Vortex''), representing an accuracy improvement of $\sim 50\,\%$.}
\label{fig:phase_exp_unwrapped}
\end{figure*}

In order to present the U-net performance more detailed we consider two scenarios with given phase ground truth $\phi_{\text{gt}}$. The corresponding measured intensity signals $I_1$, $I_2$ are depicted in Fig.~\ref{fig:phase_exp_unwrapped} without (``Gaussian'') and with spiral phase diversity (''Vortex''), see subfigures (a)\,--\,(h). The predicted phase profiles $\phi_{\text{pre}}$ (m)\,--\,(o) can be compared to the ground truth $\phi_{\text{gt}}$ (i)\,--\,(l). Additionally, the difference of both signals $\phi_{\text{diff}}$ is plotted (q)\,--\,(t) including the corresponding $\delta$-parameter. The examples discussed illustrate in detail what the statistics have already told us, cf.~Fig.~\ref{fig:loss function}. Using spiral phase diversity enables to detect smaller intensity features and thus to make more accurate predictions about the associated phase distributions where $\delta$-parameters are reduced by $\sim 30\%$ and $50\%$, respectively.

\subsection{Beam correction verification}
\label{verification}
In the previous section, we have demonstrated in experiments that predicted phase distributions meet the ground truth with deviations that have almost no optical relevance ($\delta \approx \lambda/99$). In this section, we would like to present how an adaptive optics can be controlled to compensate for aberrations and restore field distributions that are almost identical to a reference in real-time. Using the setup shown in Fig.~\ref{AOsetup} with the SLM as adaptive optical element (cf.~Sec.~\ref{AOsetup}\,\ref{setup}), again the two intensity signals $I_1$, $I_2$ are used to train the network. Here, the reference signals are obtained from a plane phase ground truth $\phi_{\text{gt}}\left(x,y\right) = 0$. \textcolor{black}{Similar to guide stars in adaptive optics for astronomy \cite{rigaut1992laser}, a reference is essential in our machine learning approach, too. The highest achievable focus quality is always determined by the reference used for the training.}

\begin{figure}
\centering
\includegraphics[width=\columnwidth]{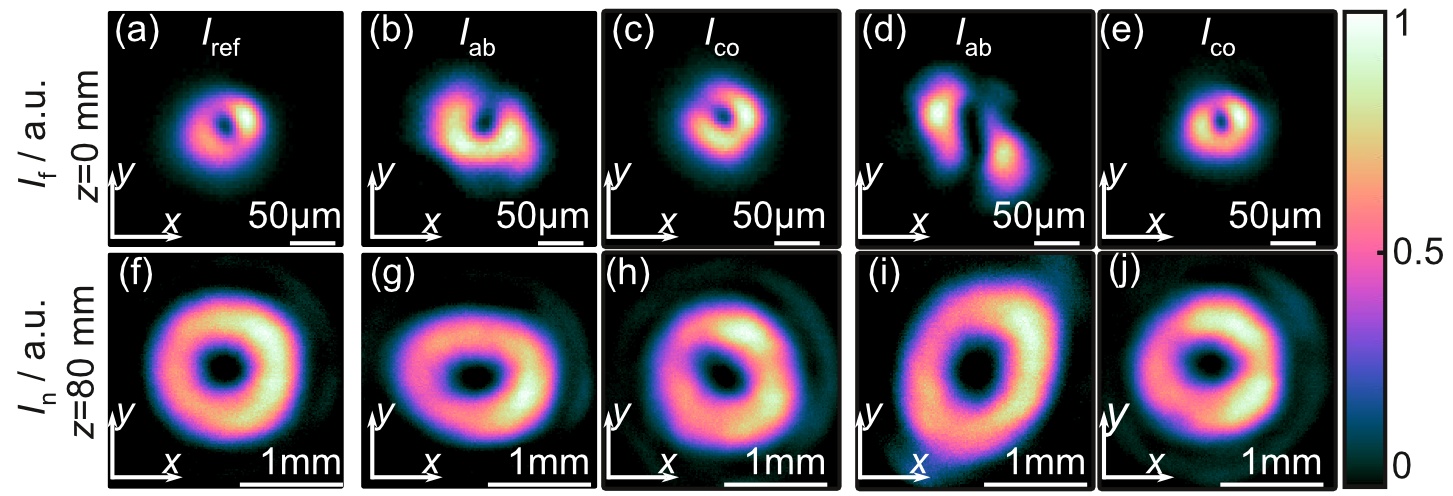}
\caption{Demonstrating the aberration correction from intensity-only predicted phase profiles. Measured far- and near-field-like reference intensity distributions $I_{\text{ref}}$ (a), (f). Measured aberrated far- and near-field-like intensity distributions $I_{\text{ab}}$ (b), (g) as well es corresponding corrected intensity distributions $I_{\text{co}}$ (c), (h). Second example of aberrated (d), (i) and corrected (e), (j) intensity signals. In all cases spiral phase diversity was applied.}
\label{fig:correct}
\end{figure}

\begin{figure}
\centering
\includegraphics[width=\columnwidth]{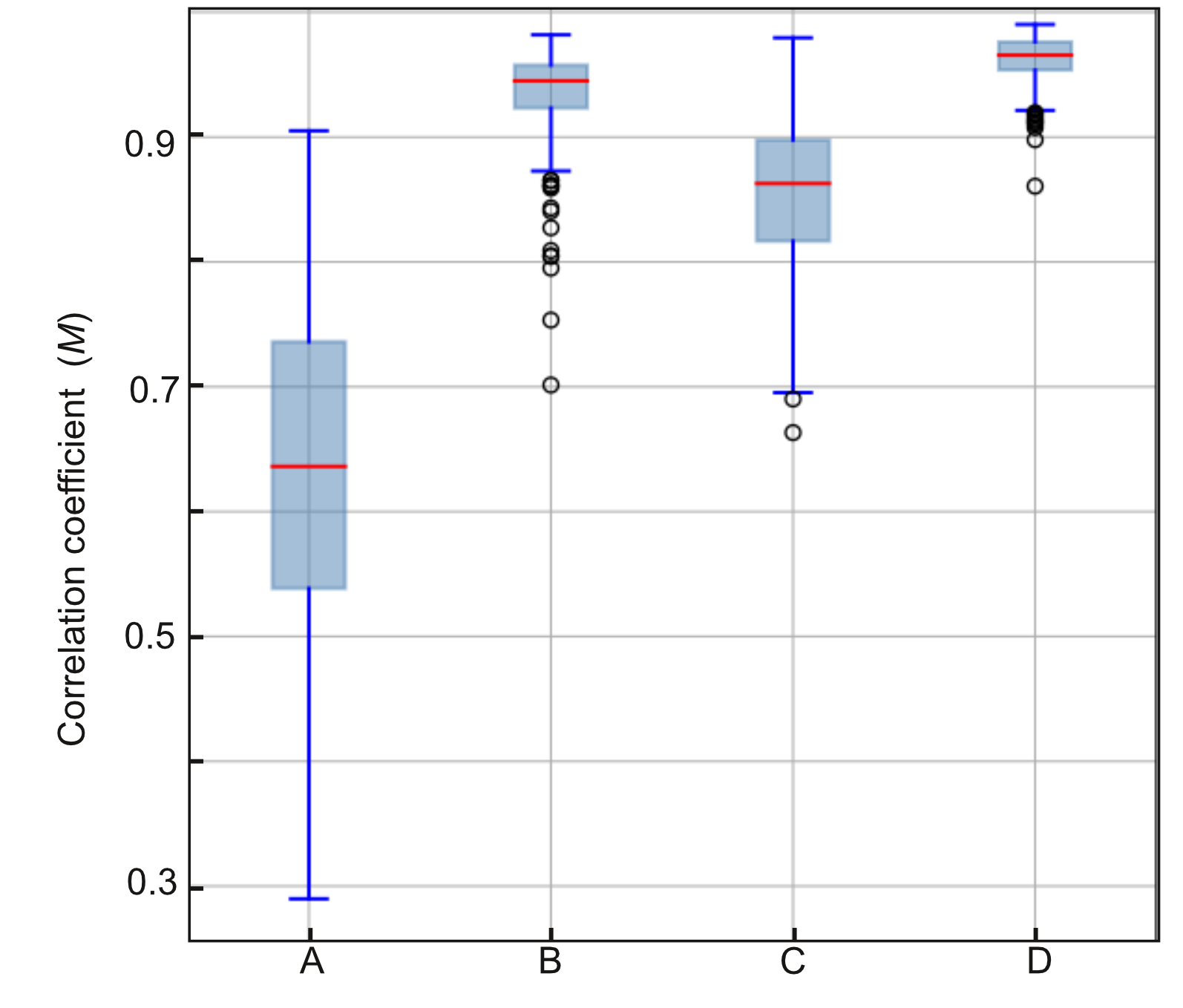}
\caption{Correlation coefficients (M) distribution for 200 aberrated and corrected fields: (A) Aberrated far field; (B) Corrected far field; (C) Aberrated near field; (D) Corrected near field. Corrected fields show higher mean M values: $\sim 0.93$(far) and $\sim 0.96$ (near), compared to aberrated fields: $\sim 0.63$ (far) and $\sim 0.86$ (near).}
\label{fig:coeff}
\end{figure}

Assuming an ideal fundamental mode operation from our single-mode-fiber-based source we expect ring-like near- and far-field intensity distributions after the SPP illumination (very close to a Laguerre-Gaussian mode of zero radial, and $\ell = 1$ azimuthal order \cite{kogelnik1966laser}). This is confirmed by the intensity signals shown in Fig.~\ref{fig:correct} (a) and (f). We can only speculate here as to where the deviations from an ideal ring profile in the intensity signal originate. Non-ideal alignment or optical components could be responsible for this. However, since this state serves as a reference in our experiment, our phase compensation approach will always lead to this associated intensity signal. This trend is visualized from two selected examples with aberrated $I_{\text{ab}}$ near- and far-field intensities, see (b), (g) and (d), (i), respectively, and corresponding compensations $I_{\text{co}}$ (c), (h) and (e), (j), respectively. 
The compensated intensity signals show a high degree of similarity to the reference. To quantify this similarity we make use of the two-dimensional cross-correlation coefficient $M$ \cite{nyberg1972optical} and evaluate experimentally 200 randomly generated cases. The statistic plotted in Fig.~\ref{fig:coeff} confirms our observation from the two selected examples discussed in Fig.~\ref{fig:correct}. After the compensation the intensity signals $I_1$, $I_2$ exhibit a higher level of similarity to our reference. The box plot in row A (aberrated far-fields) has its median at $M \approx 0.63$ which is enhanced to $M \approx 0.93$ (row B, compensated far-field). A similar trend is seen at the far-fields where the median of the correlation coefficient is enhanced from $\approx 0.86$ (row C, aberrated near-field) to $\approx 0.96$ (row D, compensated near-field).

In the present case, the prediction of the existing phase disturbances allowed us to restore the reference state in-situ by means of adaptive optics, i.e.~to achieve a flat phase and thus best possible beam quality. The time required for the prediction is $\sim \unit[17]{ms}$ on a conventional personal computer. The reference can be given to the U-Net either experimentally or by theoretical distributions---e.g. by assigning a (calculated) diffraction-limited focus and a plane phase distribution. 

The discussed technique is of particular relevance for high-power laser development. These laser systems often exhibit beam quality degradation when the thermal loads become stronger---e.g. near the maximum power or energy performance \cite{pfistner1994thermal}. At moderate power levels, on the other hand, operation at the spatial diffraction limit can often be achieved. A similar argumentation holds for high-energy pulsed lasers, where corresponding high intensities cause phase distortions from intensity-dependent refractive indices ($B$-integral) within optical components. Here, too, at moderate energies best possible beam quality can be at hand. This state can serve as reference where the U-Net can be trained with simple, intensity-only measurements. Finally, adaptive optics can be used to compensate for phase distortions occurring at higher power levels.

\section{Conclusion}
In conclusion, we have presented a fast and accurate method to predict phase distributions and enhance beam quality with deep learning and phase diversity in real-time ($\sim\unit[17]{ms}$ on a personal computer). The phase information from two intensity measurements is available zonal, thus as two-dimensional distribution and not in terms of mode coefficients. This is beneficial in cases where phase distortions cannot be described accurately with known polynomials. A thorough tolerance study was conducted considering noise and camera shifts with synthetic and experimental data which allowed us to explore the physical limits of our approach. Applying spiral phase diversity, predicted phase information met the ground truth with RMSE differences as small as $\sim \lambda/99$. However, the quality of the phase prediction is high even without the spiral phase diversity where we achieved RSME differences down to $\lambda/33$. We applied the concept to laser radiation which is representative for single-mode high-power laser cases. We were able to prove that the best beam quality reference can be restored by compensating for known phase distortions using adaptive optics.

\section*{Funding}
Supported by the Free State of Thuringia and the European Social Fund Plus (2022 FGR 0002) and the Federal Ministry of Research, Technology and Space (BMFTR, RUBIN-UKPino 03RU2U032F).

\section*{Disclosures}
JW, SB, SSR, VR, DF: TRUMPF Laser- und Systemtechnik SE (E); DB: TRUMPF Laser SE (E).

\section*{Data Availability}
Data underlying the results presented in this paper are not publicly available at this time but can be obtained from the authors on reasonable request.

\bibliography{Lib1}
\bigskip
\noindent
\bibliographyfullrefs{Lib1}
\end{document}